\documentclass[twocolumn,superscriptaddress,amsmath,aps,showpacs,showkeys,longbibliography]{revtex4-1}
\usepackage{graphicx}
\usepackage{dcolumn}
\usepackage{bm}
\usepackage{color}
\usepackage{multirow}

\begin{document}
\def\eq#1{(\ref{#1})}
\def\fig#1{Fig.~\ref{#1}}
\def\tab#1{Table~\ref{#1}}
\title{Pressure effects on the unconventional superconductivity of the noncentrosymmetric LaNiC$_2$}
\author{B. Wiendlocha}
\email{wiendlocha@fis.agh.edu.pl}
\affiliation{Faculty of Physics and Applied Computer Science, AGH University of Science and Technology, al. Mickiewicza 30, 
30-059 Krakow, Poland}
\author{R. Szcz{\c{e}}{\'s}niak}
\email{szczesni@wip.pcz.pl}
\affiliation{Institute of Physics, Jan D{\l}ugosz University in Cz{\c{e}}stochowa, Ave. Armii Krajowej 13/15, 42-200 Cz{\c{e}}stochowa, Poland}
\affiliation{Institute of Physics, Cz{\c{e}}stochowa University of Technology, Ave. Armii Krajowej 19, 42-200 Cz{\c{e}}stochowa, Poland}
\author{A. P. Durajski}
\email{adurajski@wip.pcz.pl}  
\affiliation{Institute of Physics, Cz{\c{e}}stochowa University of Technology, Ave. Armii Krajowej 19, 42-200 Cz{\c{e}}stochowa, Poland}
\author{M. Muras}
\affiliation{Faculty of Physics and Applied Computer Science, AGH University of Science and Technology, al. Mickiewicza 30, 
30-059 Krakow, Poland}
\date{\today} 
\begin{abstract}
The unconventional superconductivity in the noncentrosymmetric LaNiC$_2$, and its evolution with pressure, is analyzed basing on the 
{\it ab initio} computations and the full Eliashberg formalism. First principles calculations of the electronic structure, phonons and the electron-phonon coupling are reported in the pressure range 0-15 GPa. The thermodynamic properties of the superconducting state were determined numerically solving the Eliashberg equations. We found that already at $p=0$ GPa, the superconducting parameters deviate from the BCS-type, and a large value of the Coulomb pseudopotential $\mu^{\star}=0.22$ is required to get the critical temperature $T_c = 2.8$~K consistent with experiment. If such $\mu^{\star}$ is used, the Eliashberg formalism reproduces also the experimentally observed values of the superconducting order parameter, the electronic specific heat jump at the critical temperature, and the change of the London penetration depth with temperature.
This shows, that deviation of the above-mentioned parameters from the BCS predictions do not prejudge on the triplet or multiple gap nature of the superconductivity in this compound.
Under the external pressure, calculations predict continuous increase of the electron-phonon coupling constant in the whole pressure range 0-15~GPa, consistent with 
the experimentally observed increase in $T_c$ for the pressure range 0-4~GPa, but inconsistent with the drop of $T_c$ above 
4~GPa and the disappearance of the superconductivity above 7~GPa, reported experimentally. 
The disappearance of superconductivity may be accounted for by increasing 
the $\mu^{\star}$ to 0.36 at 7~GPa, which supports the hypothesis of the formation of a new high-pressure electronic phase, which competes with the superconductivity. 
\end{abstract}
\maketitle
%
%
\section{Introduction} 

\begin{figure}[b]
\includegraphics[width=0.45\textwidth]{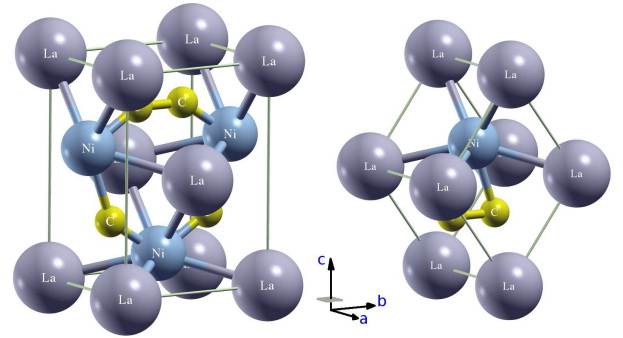}
\caption{The crystal structure of the LaNiC$_2$ compound: the base-centered conventional unit cell (left panel) 
         and the primitive cell (right panel). Graphics generated using {\sc xcrysden}~\cite{Kokalj2003A}.}
\label{fig:cryst}
\end{figure}

LaNiC$_2$ is a noncentrosymmetric compound belonging to the family of ternary nickel carbides $R$NiC$_2$, where $R$ are the rare-earth elements. It crystallizes in the base centered orthorombic CeNiC$_2$-type structure with the space group {\it Amm}2, and can be obtained using the arc metling method~\cite{Lee1996A} or the solvothermal route~\cite{Yuanzhu2007A}. The structure of LaNiC$_2$ lacks the inversion symmetry in the NiC$_2$ plane (see, Fig.~\ref{fig:cryst}). 

The low-temperature superconductivity in LaNiC$_2$ was first reported by Lee {\it et al.}~\cite{Lee1996A} with the critical temperature 
$T_c=2.7$~K. In the later papers, Lee {\it et al.}~\cite{Lee1997A, Lee1997B} showed that substituting lanthanum atoms with thorium increases the critical temperature up to 7.9~K, {whereas nitrogen substitution for carbon may increase $T_c$ to about 4 K, as reported by Syu {\it et al.}~\cite{Syu2010A}}. The noncentrosymmetric superconducting materials attracted wider attention since the discovery of the heavy fermion superconductor CePt$_3$Si~\cite{Bauer2004A}. Lack of the inversion symmetry, if accompanied by the spin-orbit coupling, leads to the splitting of the electronic bands and may lead to the mixing of the spin-singlet and the triplet pairing states. Only a few such systems were reported up-to-date. Among them there is a group of materials with the strong electron-electron interactions, containing already mentioned CePt$_3$Si, as well as UIr~\cite{Akazawa2004A}, CeRhSi$_3$~\cite{Kimura2007A} or CeCoGe$_3$~\cite{Masson2009A}. 
In these systems, superconductivity appears in the antiferromagnetic ordered phase or around the magnetic quantum critical points~\cite{Katano2014A}. We may also distinguish the second group of the 
noncentrosymmetric superconductors, with weak electron correlations, where one can find Li$_2$(Pd,Pt)$_3$B~\cite{Togano2004A, Nishiyama2007A} or Mg$_{10}$Ir$_{19}$B$_{16}$~\cite{Klimczuk2006A}. 
The LaNiC$_2$ compound seems to be located somewhere between these two groups of strongly correlated and not correlated materials.

Up to now there is no general agreement, whether LaNiC$_2$ should be categorized as a conventional, $s$-wave BCS-like superconductor, or it should be considered as unconventional, with multiple superconducting gaps or even the spin-triplet pairing. 

The thermodynamic properties of the superconducting state of LaNiC$_2$ were experimentally investigated with rather high accuracy. The various parameters, characterizing the superconducting state are collected in \tab{tab:1}. Lee {\it et al.}~\cite{Lee1996A} reported anomalous temperature dependence of the specific heat in the superconducting state. The normal state specific heat $C^{N}$ was fitted as typical: 
$C^{N}\left(T\right)=\gamma T+\beta T^{3}$ (3~K $\leq$ $T$ $\leq$ 7~K), where $\gamma=7.83$ ${\rm mJ/mol K^{2}}$ and $\beta=0.0635$ ${\rm mJ/mol  K^{4}}$. On the other hand, the specific heat in the superconducting state ($C^{S}$) was not exponential but had the power-law form: 
$C^{S}\left(T\right)\simeq 3.5\left(\gamma T_c\right)\left(T/T_c\right)^{3}$, in the temperature range 0.35~K $\leq$ $T$ $\leq$ 1.6~K. 
Basing on this non-exponential decay of the specific heat below $T_c$, the authors concluded that LaNiC$_2$ was an unconventional superconductor with dominant $p$-wave symmetry of the order parameter~\cite{Lee1996A}. The normal-state value of $\gamma$ was confirmed by 
Chen {\it et al.}~\cite{Chen2013A}, where $\gamma=7.7$ ${\rm mJ/mol  K^{2}}$, and the observed temperature dependence of the specific heat was analyzed assuming that LaNiC$_2$ was a two-gap BCS superconductor~\cite{Chen2013A}, in analogy to $\rm MgB_{2}$~\cite{Nagamatsu2001A, Szczesniak2008A}.

\begin{table}[t]
\centering
\caption{The experimental values of the thermodynamic parameters obtained for the LaNiC$_2$ superconductor. 
$R_{\Delta}= 2\Delta(0)/k_{B}T_c$, $R_{C}=\Delta C\left(T_c\right)/\gamma T_c$, 
where $\Delta(0)$ is the superconducting energy gap at $T = 0$~K, and  $\Delta C\left(T_c\right)$ is the jump of the electronic specific heat at the critical temperature.
}
\label{tab:1}
\begin{ruledtabular}
\begin{tabular}{lllllll} 
$p$ (GPa)                                            & 0          & 2        & 4        &  6        & $\sim 7$  &    Ref. \\ 
                                                      \hline
                                                     & 2.8        & 3.4      & 3.8      & 2. 5      & $\sim 0$  &  \cite{Katano2014A}     \\
$T_c$ (K)                                            & 2.7-2.9    & -        & -        &  -        & -        &  \cite{Lee1997B}        \\            
                                                     & 2.7-2.75   & -        & -        &  -        & -        &  \cite{Hirose2012A}     \\                                                   
                                                     & 3.01-3.25  & -        & -        &  -        & -        &  \cite{Syu2010A}        \\                                                                                                         
\hline
                                                     & 2.5\footnote{The value found based on a fit of the  
                                                     experimental data (the penetration depth at low temperatures) to the one-gap model. In the original paper \cite{Chen2013A} two-gap model was used.}
                                                                  & -        & -        & -         & -        &  \cite{Chen2013A}       \\                                                                                                               
$R_{\Delta}$                                         & 2.9        & -        & -        & -         & -        &  \cite{Hirose2012A}     \\
                                                     & 3.34       & -        & -        & -         & -        &  \cite{Iwamoto1998A}    \\ 
\hline   
                                                     & 1.2        & -        & -        & -         & -        &  \cite{Lee1996A}        \\
$R_{C}$                                              & 1.05       & -        & -        & -         & -        &  \cite{Chen2013A}       \\  
                                                     & 1.26       & -        & -        & -         & -        &  \cite{Pecharsky1998A}  \\ 
\end{tabular}
\end{ruledtabular} 
\end{table} 

Contrary to that, Pecharsky {\it et al.}~\cite{Pecharsky1998A} reported much lower $\gamma = 6.5(2)$ ${\rm mJ/mol K^{2}}$ and observed an exponential decay of $C^{S}$ (over the temperature range 1.52~K $\leq$ $T$ $\leq$ $T_c$), which together with the nuclear quadrupole resonance
study by Iwamoto {\it et al.}~\cite{Iwamoto1998A} supported the conventional BCS-type of the superconductivity. The further studies by Hirose 
{\it et al.} \cite{Hirose2012A} of the low temperature specific heat supported this claim.

London magnetic penetration depth $\lambda_L$ and the muon spin-relaxation ($\mu$SR) measurements gave independent hints for the unconventional character of the superconductivity in LaNiC$_2$. The two-gap BCS superconductivity was proposed basing on $\lambda_L(T)$ measurements by Chen {\it et al.}~\cite{Chen2013A}. On the other hand, Bonalde {\it et al}~\cite{Bonalde2011A} reported magnetic penetration 
depth characteristics, that suggested existence of nodes in the energy gap. Hillier {\it et al.}~\cite{Hillier2009A} conducted $\mu$SR measurements that detected spontaneous magnetic fields which indicated that the time reversal symmetry was broken in the superconducting state. 
The further analysis by Quintanilla {\it et al.}~\cite{Quintanilla2010A} concluded that this result could only be compatible with the non-unitary triplet pairing states, where the superconducting instability must have been split by the spin-orbit coupling (SOC). 
Interestingly, SOC itself had to be relatively small to allow for such a state~\cite{Quintanilla2010A}. The time-reversal symmetry breaking has been also recently reported by Symiyama {\it et al.}~\cite{Sumiyama2015A}, where a spontaneous magnetization on the order of 10$^{-5}$~G along c-axis has been found.

The reduced experimental values of the superconducting energy gap $2\Delta\left(0\right)$, and the specific heat jump at the critical temperature  
$\Delta C\left(T_c\right)$ are collected in Table~\ref{tab:1}. Not determining the type of the pairing symmetry, the obtained results suggest 
that LaNiC$_2$ belongs to a group of the superconductors with relatively weak coupling. However, the values of the ratios: 
$R_{\Delta}= 2\Delta\left(0\right)/k_{B}T_c$ and $R_{C}=\Delta C\left(T_c\right)/C^{N}\left(T_c\right)$, ($C^{N}$ is the specific heat in the normal state) are below these, predicted by the BCS theory: $\left[R_{\Delta}\right]_{\rm BCS}=3.53$ and $\left[R_{C}\right]_{\rm BCS}=1.43$~ \cite{Bardeen1957A, Bardeen1957B}, which is not the common case for the electron-phonon superconductors~\cite{Carbotte1990A}. 

\begin{table}[b]
\caption{Theoretical crystal structure parameters for LaNiC$_2$, after relaxation under various pressures, in the orhorombic base-centered unit cell {\it Amm}2. The atomic positions are as follows: La $(0,0,u)$, Ni $(0.5,0,v)$ and C $(0.5,\pm y,z)$. Experimental values are given for comparison.}
\label{tab:relax}
\begin{center}
\begin{ruledtabular}
\begin{tabular}{ l c c c c c c c}
P 	    &   	$a$&	$b$&	$c$&	      $u$&	$v$&	$y$&	$z$\\
(GPa)	&   	(\AA)&	(\AA)&	(\AA)&	         &	   &	   &	   \\
\hline
0\footnotemark[1] &       3.952     &4.557       &6.193  &0.0    &       0.626 & 0.160 & 0.289\\
0	&        3.987	   &4.552	&6.166	&-0.0048&	0.6098&	0.1494&	0.2995\\
2	&        3.956	   &4.539	&6.140  &-0.0048&	0.6096&	0.1498&	0.2996\\
4	&        3.929	   &4.527	&6.116	&-0.0047&	0.6093&	0.1502&	0.2997\\
5.5	&        3.908     &4.516 	&6.101	&  -0.0046&	0.6090&	0.1505&	0.2998\\
7	&        3.889	   &4.508	&6.085	&-0.0045&	0.6088&	0.1508&	0.2998\\
10	&        3.853 	   &4.490	&6.057	& -0.0044&	0.6085&	0.1513&	0.2999\\
15	&        3.799 	   &4.463	&6.015	& -0.0042&	0.6080&	0.1521&	0.3001
\end{tabular}
\footnotetext[1]{Experimental values, Ref.~\cite{Katano2014A}}
\end{ruledtabular}
\end{center}
\end{table}

Recently, another unconventional behavior of LaNiC$_2$ has been discovered by Katano {\it et al.}~\cite{Katano2014A}, who reported anomalous evolution of $T_c$ with pressure (see also Table~\ref{tab:1}). At first, the external pressure increases $T_c$, from 2.8 K at $p=0$~GPa to $\sim 3.8$~K, fairly constant between 3 and 4~GPa. This itself is a rare situation among the conventional superconductors, where usually $T_c$ decreases due to the crystal lattice stiffening~\cite{Wiendlocha2015}. For larger pressures, $T_c$ drops down, and LaNiC$_2$ was reported not to superconduct above 
$\sim 7-8$~GPa \footnote{Ref.~\cite{Katano2014A} does not give the precise value of the critical pressure where the superconductivity disappears, in our computations we consistently took 7 GPa as the critial value.}. Basing also on an analysis of the normal state resistivity, the authors suggested~\cite{Katano2014A} that 
for $p>4$~GPa different, the high-energy-scale correlated electronic state emerges, and the characteristic temperature of this unidentified phase increases with pressure, from $\sim 40$~K ($p=4$~GPa) to $\sim 240$~K ($p=8$~GPa). As possible candidates for this new phase, the charge density wave (CDW) or the Kondo phases were mentioned~\cite{Katano2014A}, but the nature of this phase remained unclassified. Note, that in many other compounds of the 
$R$NiC$_2$-type, CDW were observed at high temperatures~\cite{Murase2004A}. However at high pressures the CDW state is usually not stable~\cite{Katano2014A}.

The aim of this work is to theoretically investigate the effect of pressure on the electronic structure, phonons and the superconductivity in LaNiC$_2$, assuming that the superconductivity is mediated by the electron-phonon interaction. In the first step, the {\it ab initio} calculations of the band structure and the electron-phonon coupling in the pressure range 0-15~GPa were carried out. Next, the thermodynamic parameters of the superconducting state were determined using the Eliashberg formalism~\cite{Eliashberg1960A}, where the depairing effects were parameterized~\cite{Carbotte1990A, Morel1962A}. The main problems and the questions which are raised in this work are: 
(i) how the external pressure changes the electronic structure and dynamical properties of the system; 
(ii) may the electron-phonon interaction be responsible for the observed increase in $T_c$ under pressure; 
(iii) may the electron-phonon interaction be responsible for the disappearance of the superconductivity above 7~GPa;
(iv) is the superconductivity in LaNiC$_2$ at $p = 0$~GPa unconventional in view of the Eliashberg formalism? 

%
\section{{\it Ab initio} computations}\label{sec:abinitio}

First principles calculations of the electronic structure, phonons and the electron-phonon interaction were performed using two methods: 
the full potential linearized augmented plane wave method (FP-LAPW, {\sc WIEN2k} code~\cite{Blaha2001A}) and the plane-wave pseudopotential method ({\sc Quantum ESPRESSO}, QE package~\cite{Giannozzi2009A}). The relaxed crystal structures,  phonons and the electron-phonon interaction functions under various pressures in the range from 0 to 15~GPa were computed using QE package in the scalar-relativistic approximation, {thus, the spin-orbit coupling is not taken into account in the duscussion of superconducting phase}. The projector augmented wave (PAW) pseudopotentials~\cite{pseudo} were employed, the plane-wave and the charge density cut-offs were set to 70~Ry and 600~Ry, respectively. A (4,4,4) {\bf q}-point mesh (i.e. 21 different  {\bf q}-points) was used for the calculations of the dynamical matrices and other related phonon properties in the reciprocal space, with a (16,16,16) {\bf k}-point Brillouin zone sampling.  

The electronic structure and the spin-orbit coupling (SOC) effects are analyzed basing on LAPW computations results, carried out on a very fine mesh of about 60 000 {\bf k}-points. Perdew-Burke-Ernzerhof Generalized Gradient Approximation~\cite{Perdew1996A} (PBE-GGA) exchange-correlation 
potential was used in all the computations. 

The obtained crystal structure parameters in the {\it Amm}2 base-centered orthorombic unit cell are gathered 
in Table~\ref{tab:relax}. The ambient pressure values are in very good agreement with the experimental ones~\cite{Katano2014A}, and the theoretical unit cell volume is only 0.3\% larger. When the pressure increases, the strongest reduction in the unit cell dimensions is seen along $a$ axis, i.e. perpendicular to the Ni-C and C-C bonds, see Fig.~\ref{fig:cryst}. For the largest studied pressure of $p = 15$~GPa, the compressions of the unit cell edges are about 5\%, 2\% and 2.5\% for $a$, $b$ and $c$, respectively. 

\begin{table}[t]
\caption{Densities of states at the Fermi level for the selected pressures (SOC included in the calculations). 
         The partial atomic densities are projected on atomic spheres with the radius: 2.5 a.u. (La), 2.05 a.u. (Ni), and 1.25 a.u. (C).}
\label{tab:dos}
\begin{center}
\begin{ruledtabular}
\begin{tabular}{ l c c c c c c c}
P 	    &   	$N(E_F)$&	Ni&  Ni-3d&La     &La-5d  &C      &C-2p\\
(GPa)	&   	(eV$^{-1}$)&   (eV$^{-1}$)&(eV$^{-1}$)&(eV$^{-1}$)&(eV$^{-1}$)&(eV$^{-1}$)&(eV$^{-1}$)\\
\hline
0	&        2.37	   &0.57	&0.49	& 0.47&	0.34&	0.10&	0.09\\
4	&        2.28	   &0.60	&0.53	& 0.44&	0.32&	0.09&	0.08\\
7	&        2.27	   &0.60	&0.52	& 0.44&	0.32&	0.09&	0.08\\
15	&        2.33 	   &0.63	&0.56	& 0.45&	0.33&	0.10&	0.09
\end{tabular}
\end{ruledtabular}
\end{center}
\end{table}
%

%
\subsection{Electronic structure}
\begin{figure}[t]
\includegraphics[width=0.48\textwidth]{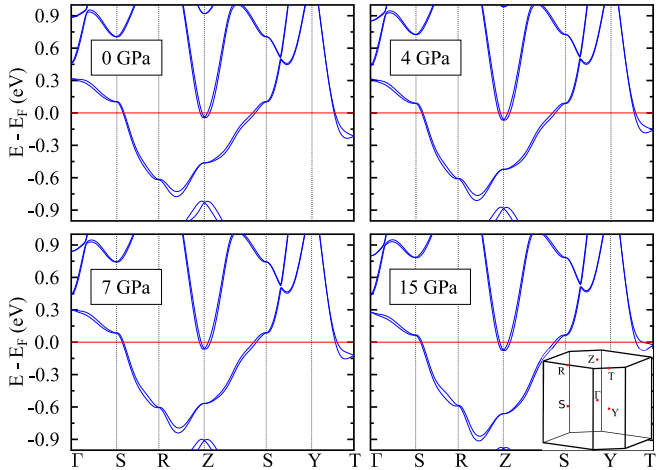}
\caption{Electronic dispersion relations near $E_F$ for LaNiC$_2$, in the pressure range form 0 to 15~GPa. The inset shows the Brillouin zone with the 
         high symmetry points. Band splitting due to SOC is well visible.}
\label{fig:bands}
\end{figure}
\begin{figure}[t]
\includegraphics[width=0.48\textwidth]{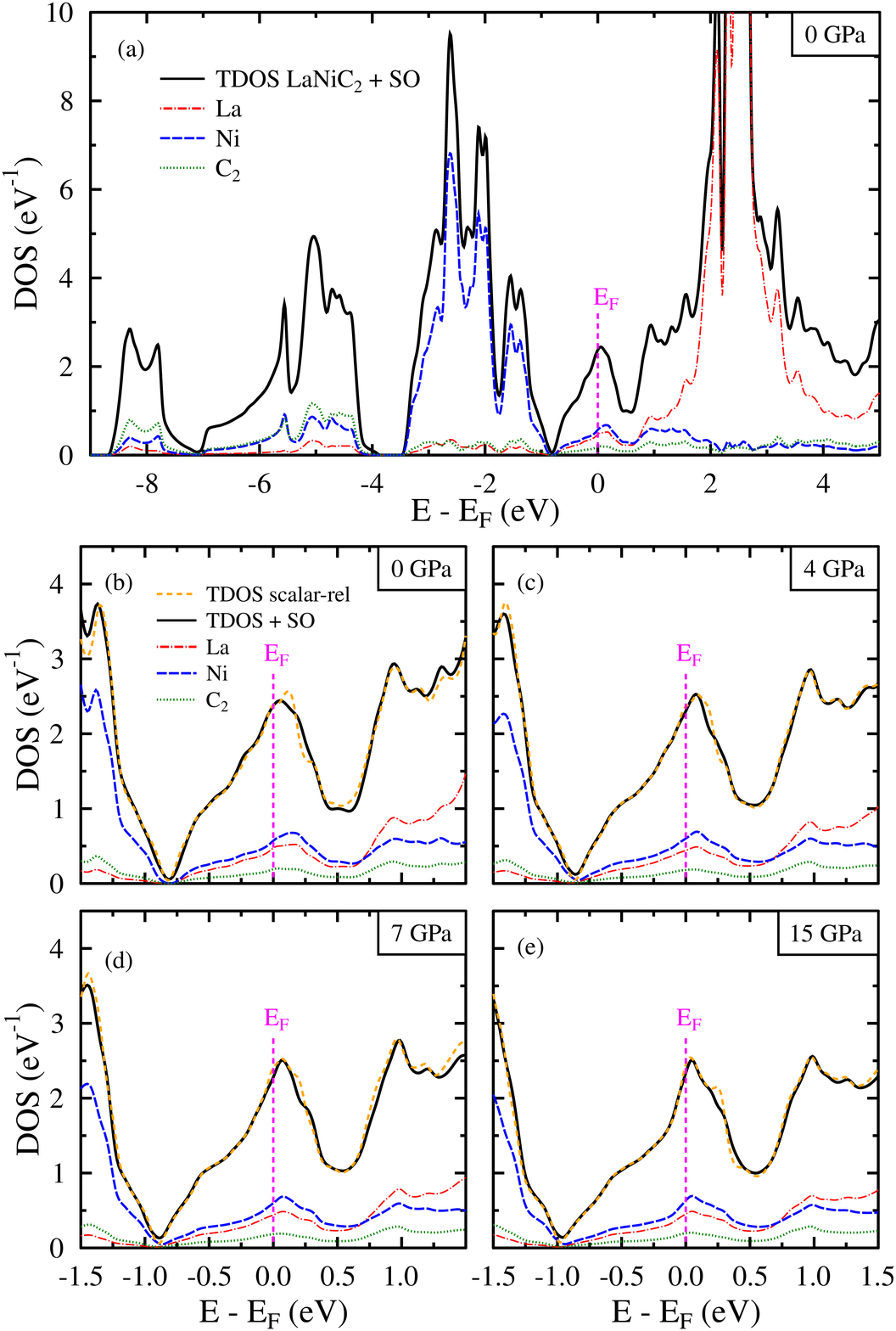}
\caption{Electronic densities of states (DOS) for LaNiC$_2$. Panel (a): Total DOS (TDOS) and the partial atomic densities, plotted in a broad 
         energy range, for $p = 0$~GPa. Panels (b-e): DOS near $E_F$ for selected pressures. TDOS from scalar-relativistic computations, i.e. 
         neglecting SOC, added for comparison (orange dashed line).}
\label{fig:dos}
\end{figure}

\begin{figure*}[t]
\includegraphics[width=\textwidth]{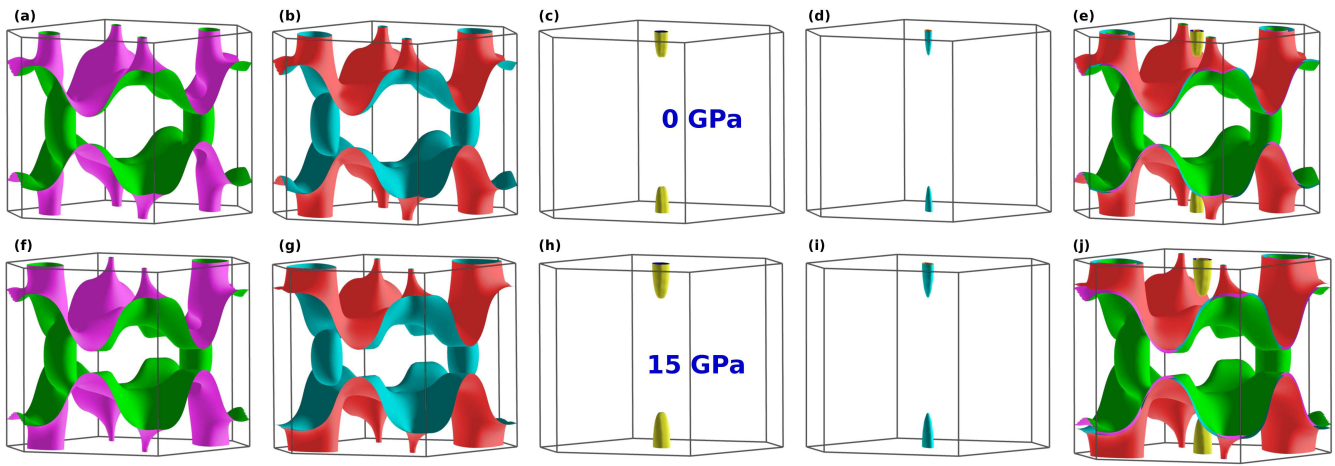}
\caption{The Fermi surface of LaNiC$_2$ for 0~GPa (upper panel) and 15~GPa (lower panel). Panels (a-d) and (f-i) show each of the FS sheets 
         separately, whereas panels (e) and (j) show the total Fermi surface.}
\label{fig:FS}
\end{figure*}

The electronic structure of LaNiC$_2$ was studied under ambient conditions theoretically~\cite{Laverock2009A, Subedi2009A, Hase2009A} and experimentally~\cite{Hirose2012A}. First principles calculations by Subedi and Singh~\cite{Subedi2009A} suggested that LaNiC$_2$ is a conventional, non-magnetic electron-phonon superconductor, with the intermediate coupling, where the large contributions to the electron-phonon coupling comes from two low-frequency C nonbond-stretching phonon modes. Our results for the selected four pressures are presented 
in Fig.~\ref{fig:bands} (dispersion relations), Fig.~\ref{fig:dos} (densities of states, DOS) and Fig.~\ref{fig:FS} (Fermi surfaces, FS).
For $p = 0$~GPa our results are generally consistent with above-mentioned works. The Fermi level crosses four bands, which arise from two bands split by the spin-orbit interaction in the absence of the center of inversion in the unit cell. This leads to four groups of the sheets in the Fermi surface pictures (Fig.~\ref{fig:FS}). FS consist of two large and complex hole-like sheets [Fig.~\ref{fig:FS}(a-b)], and two sets of two very small ellipsoidal electron pockets centered around Z point in the Brillouin zone (BZ) [Fig.~\ref{fig:FS}(c-d)]. Fig.~\ref{fig:FS}(e) shows all the sheets plotted together. The electronic states near $E_F$, that build the Fermi surface, are mainly due to the 3d-Ni and 5d-La orbitals, with smaller contribution from the carbon atoms, which is supported by the partial atomic densities of states plotted in Fig.~\ref{fig:dos}. The values of DOS at $E_F$ for several pressures are presented in Table~\ref{tab:dos}. The values of the partial densities result from projections onto atomic 
spheres with the 
radius: 2.5 a.u. (La), 2.05 a.u. (Ni) and 1.25 a.u. (C), kept constant while changing pressure. 
For all the pressures, the ground state is not magnetic in calculations. Computed value of $N(E_F)$ for $p = 0$~GPa allows to estimate the magnitude of the electron-phonon coupling, assuming that the electronic specific heat is renormalized only by the electron-phonon interaction: 
$\gamma_{\rm expt.}=\gamma_{\rm calc.}(1+\lambda)$, where $\lambda$ is the electron-phonon interaction parameter. 
Taking the experimental values from Refs.~\cite{Lee1996A, Chen2013A} 
($\gamma_{\rm expt.} \simeq 7.7-7.8$ ${\rm mJ/mol K^{2}}$) one gets $\lambda \sim 0.4$, confirming the weak coupling regime. 
The spin-orbit interaction, although very important in the context of the possible superconducting pairing symmetry, does not have a strong impact on the larger-energy-scale features of the electronic structure of LaNiC$_2$. In Fig.~\ref{fig:dos}(b), where we have DOS near $E_F$ plotted for $p=0$~GPa, the total DOS resulting from the scalar-relativistic computations has been additionally plotted. 
As one can see, no visible differences are present at or below $E_F$, and the same conclusion holds when the pressure increases, as presented in Figs.~\ref{fig:dos}(c-e). The SO splitting of the bands, as they cross $E_F$ in Fig.~\ref{fig:bands}(a) are about: 30~meV (crossing in the S-R direction), 35~meV (crossings near the Z point), 30~meV (crossing in the Z-S direction) and 35~meV (crossing in the Y-T direction), close to the average SO splitting of 3.1~mRy = 42~meV reported by Hase and Yanagisawa~\cite{Hase2009A}.
{Magnitude of the SO splitting of bands is similar to that observed e.g. in noncentrosymmetric superconductor Li$_2$Pd$_3$B (up to 30 meV)~\cite{Pickett2005}, where single-gap s-wave pairing symmetry was proposed~\cite{Yuan2006,Nishiyama2007A}, or to La$_2$C$_3$ (20-30 meV)~\cite{Kim2007}, where both single-gap s-wave~\cite{Kim2007} or nodeless two-gap~\cite{Kuroiwa2008} superconductivity were suggested.
Among weakly-correlated noncentrosymmetric superconductors, the largest SOC splitting (200 meV)~\cite{Pickett2005} was detected in spin-triplet~\cite{Yuan2006,Nishiyama2007A} superconductor Li$_2$Pt$_3$B (see, also recent review~\cite{Smidman2016} and references therein).
}

In the studied pressure range of 0-15~GPa, the electronic structure changes slightly. In the dispersion relations plots in Fig.~\ref{fig:bands}, one can identify two opposite trends in band shifts, while increasing pressure. The hole-like band near S and the electron-like near Z points go down, relative to $E_F$, whereas the electron-like part of the band near T goes up. Nevertheless, to some extent the effect of the external pressure is comparable to the effect of the electron doping, since generally $E_F$ moves towards higher-energy states, and the electron-like pockets in Fig.~\ref{fig:FS}(h-i) become larger, than in Fig.~\ref{fig:FS}(c-d). This trend is seen in DOS plots as well: 
in Fig.~\ref{fig:dos}(b-e) the local maximum of DOS, which is above $E_F$, comes closer to $E_F$ when the pressure increases. The $N(E_F)$ value, however, does not change monotonically (Table~\ref{tab:dos}), since it decreases slightly for smaller pressures, and increases for $p=15$~GPa. 
The small, continuous increase of DOS is observed on the Ni atoms only. It is worth noting that the effects of the electron doping and the external pressure on superconductivity of LaNiC$_2$ are correlated also in the experimental studies. In the electron-doped (Th, N-substituted) LaNiC$_2$~\cite{Lee1997B,Syu2010A} $T_c$ was found to increase, similarly as in non-doped LaNiC$_2$ in the pressure range 0-4~GPa.
As far as the spin-orbit effects are concerned, the SO splittings of bands that cross $E_F$ in Fig.~\ref{fig:bands}(d) for $p=15$~GPa are: 26, 47, 29 and 68~meV, in the same order as given above for $p = 0$. Thus, they increase in some parts of BZ, but decrease in other, showing no general trend.

%
\subsection{Phonons and the electron-phonon coupling}

\begin{figure*}[t]
\includegraphics[width=0.75\textwidth]{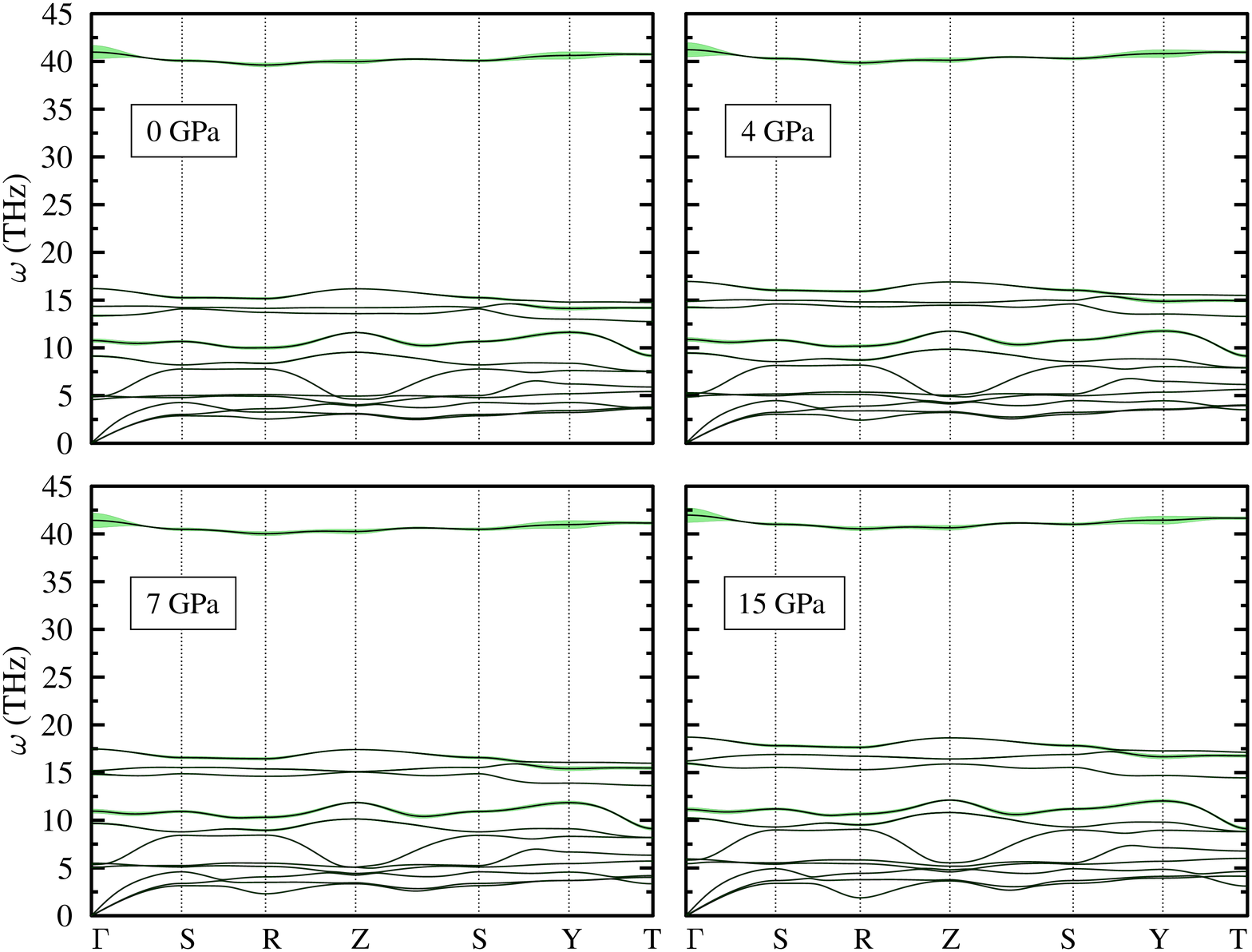}
\caption{The phonon dispersion relations for LaNiC$_2$ under four selected pressures, with the phonon linewidth marked.}
\label{fig:phdisp}
\end{figure*}

\begin{figure*}[t]
\includegraphics[width=0.75\textwidth]{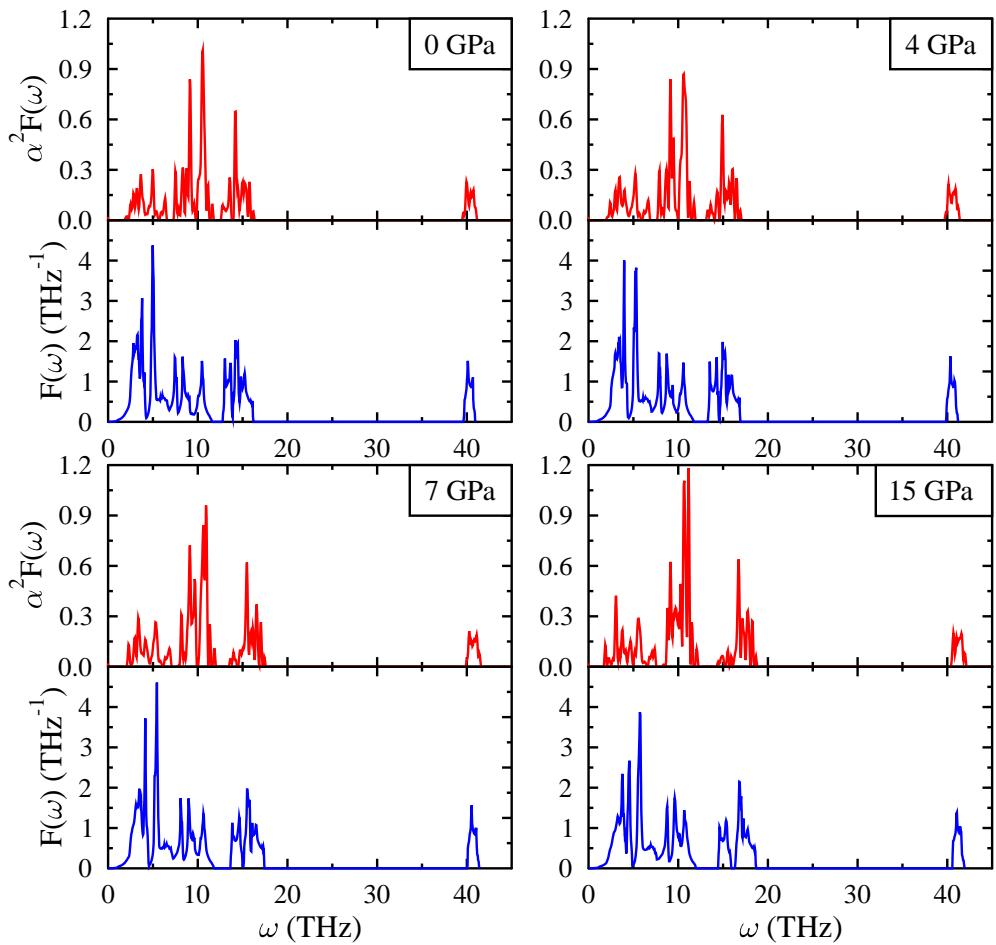}
\caption{The Eliashberg function $\alpha^2F(\omega)$ (upper panels) plotted above the phonon density of states (lower panels) for selected  
         pressures.}
\label{fig:a2f}
\end{figure*}

Now let us proceed to the analysis of the dynamical properties of LaNiC$_2$ as a function of the pressure. Figure~\ref{fig:phdisp} displays the phonon dispersion curves, with shading corresponding to the phonon linewidth $\gamma_{{\bf q},\nu}$ for the mode $\nu$ at ${\bf q}$-point, which results from the electron-phonon interaction and determine the electron-phonon coupling constant:
\begin{equation}
\lambda = \sum_{{\bf q},\nu}\frac{\gamma_{{\bf q},\nu}}{\pi \hbar N(E_F)\omega^2_{{\bf q},\nu}}.
\end{equation}
One thing that immediately catches the eye is the lonesome mode near 40~THz, which is attributed to the two carbon atoms vibrations, in which they are oscillating in opposite directions. This mode also has the largest linewidth, however, since it has a high frequency it does not give a dominating contribution to the electron-phonon coupling (see, below). This C-C bond-stretching mode was earlier identified by Subedi and Singh~\cite{Subedi2009A}, and our $p=0$~GPa results are close to theirs~\footnote{Note, that in Ref.~\cite{Subedi2009A} most likely there is a problem with the high symmetry points labels or coordinates, since all of their phonon modes at T point have exactly the same frequency as those in $\Gamma$, including the three acoustic modes with $\omega = 0$, but their phonon DOS is similar to one, reported here.}. 
Such a structure of the phonon dispersion relations divides the phonon DOS $F(\omega)$, presented in lower panels of Fig.~\ref{fig:a2f}, into two regions: below 20~THz, where 11 of the 12 phonon branches are located, and above 40~THz, where the last 12th phonon mode appears.

When external pressure is applied, the phonon frequencies are generally increasing, which is visible in the shifts of the dispersion relations, as well as in the phonon densities of states. Phonon branches grouped around 15 THz are visibly shifted with pressure, separating the phonon spectrum into three regions at larger $p$. The only phonon mode, which softens with pressure, is the first acoustic mode, especially near R and T points (see, Fig.~\ref{fig:phdisp}), but no evidence for the instability of this phonon branch was detected. Table~\ref{tab:freq} contains various averages regarding the phonon structure, and all the mean frequencies increase with pressure. They are defined as: 
\begin{table}[b]
\caption{The phonon frequency moments for the LaNiC$_2$ compound.}
\label{tab:freq}
\begin{ruledtabular}
\begin{tabular}{ c | c c c c  c}
p &	$\langle \omega^1 \rangle$&	$\langle \omega^2 \rangle$&
	$\langle \omega \rangle$& $\langle \omega_{\rm log} \rangle$ &	 $\langle \omega_{\rm log}^{\alpha^2F}\rangle $\\
(GPa) &	(THz) &	(THz) &	(THz$^2$)& (THz) & (THz)\\
	\hline
0	&6.09	&65.45	&10.75 &4.83 &6.90\\
2	&6.21	&67.79	&10.91 &4.94 &7.03\\
4	&6.35	&70.33	&11.08 &5.05 &7.08\\
5.5	&6.46	&72.59	&11.24 &5.12 &7.11\\
7	&6.51	&73.67	&11.31 &5.18 &7.13\\
10	&6.54	&74.45	&11.39 &5.19 &7.12\\
15	&6.87	&81.41	&11.84 &5.45 &6.98\\
\end{tabular}
\end{ruledtabular}
\end{table}
\begin{equation}
\langle \omega^n \rangle = \int_0^{\omega_{\mathsf{max}}} \omega^{n-1} F(\omega) d\omega / \int_0^{\omega_{\mathsf{max}}} \omega^{-1} F(\omega) d\omega,
\end{equation}
\begin{equation}
\langle \omega \rangle = \int_0^{\omega_{\mathsf{max}}} \omega F(\omega) d\omega / \int_0^{\omega_{\mathsf{max}}} F(\omega) d\omega,
\end{equation}
\begin{equation}\label{eq:omlog}
\langle\omega_{\rm log}\rangle = \exp\left(\int_0^{\omega_{\mathsf{max}}} F(\omega) \ln\omega\frac{d\omega}{{\omega}} \left/ \int_0^{\omega_{\mathsf{max}}} 
{F(\omega)}\frac{d\omega}{{\omega}} \right. \right),
\end{equation}
and
\begin{equation}\label{eq:omlog2}
\langle\omega_{\rm log}^{\alpha^2F}\rangle = \exp\left(\int_0^{\omega_{\mathsf{max}}} \alpha^2F(\omega) \ln\omega\frac{d\omega}{{\omega}} \left/ \int_0^{\omega_{\mathsf{max}}} 
{\alpha^2F(\omega)}\frac{d\omega}{{\omega}} \right. \right).
\end{equation}

The last value, $\langle\omega_{\rm log}^{\alpha^2F}\rangle$ in Table \ref{tab:freq}, is calculated from the Eliashberg coupling function 
$\alpha^2F\left(\omega\right)$, but it was added for convenience here, and its non-monotonic behavior with pressure is due to the nature of the electron-phonon coupling, which will be discussed next, not due to the changes in the lattice vibration frequencies. This logarithmic average enters the Allen-Dynes formula~\cite{Allen1975A} for the critical temperature:
\begin{equation}\label{eq:tc}
k_{B}T_{c}=\frac{\hbar\langle\omega_{\rm log}^{\alpha^2F}\rangle}{1.20}\,
\exp\left\{-\frac{1.04(1+\lambda)}{\lambda-\mu^{\star}(1+0.62\lambda)}\right\}.
\end{equation}
It is worth noting, that for LaNiC$_2$ $\langle\omega_{\rm log}^{\alpha^2F}\rangle$ it is considerably larger (40\%) than $\langle\omega_{\rm log}\rangle$, which is computed using the approximation of the frequency-independent electron-phonon coupling $\alpha^2 \simeq {\rm const.}$, equivalent to using the bare phonon DOS $F(\omega)$ function instead of the $\alpha^2F(\omega)$ for its determination, as in Eq.~\ref{eq:omlog}.

The Eliashberg coupling function is plotted against the phonon density of states in upper panels in Fig.~\ref{fig:a2f}. Comparing to the bare phonon DOS, the electron-phonon coupling is visibly enhanced near 10~THz, where also in the dispersion plots 
in Fig.~\ref{fig:phdisp} the phonon branches (7th and 8th) have larger linewidth. These modes also have the strong carbon oscillations contribution, but Ni and La atoms oscillate as well. When the pressure increases, the increase of $\alpha^2F(\omega)$ in this frequency range is observed.

\begin{figure}[t]
\includegraphics[width=0.45\textwidth]{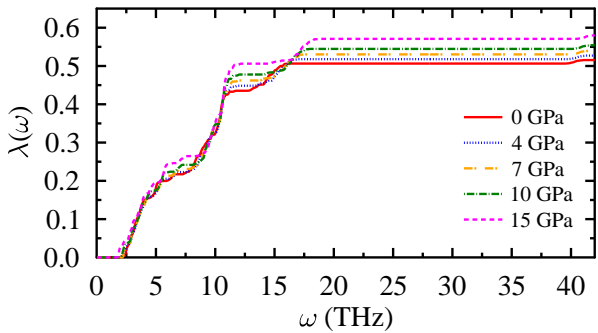}
\caption{The cummulative frequency distribution of $\lambda$, defined as: $\lambda(\omega) = 2\int_0^{\omega} \alpha^2F(\Omega) {d\Omega\over \Omega}$, 
         for selected pressures.}
\label{fig:lomega}
\end{figure}

\begin{figure}[t]
\includegraphics[width=0.48\textwidth]{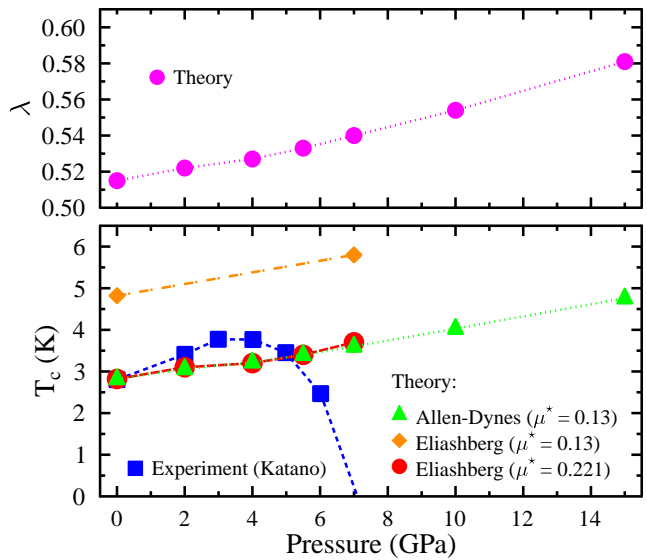}
\caption{The evolution of the electron-phonon coupling constant (upper panel) and the critical temperature (lower panel) with pressure. 
         In the lower panel: triangles: $T_c$ computed using the Allen-Dynes formula (Eq.~\ref{eq:tc}) and $\mu^{\star} = 0.13$; diamonds 
         (circles): $T_c$ computed using the Eliashberg equations and $\mu^{\star} = 0.13$ ($\mu^{\star} = 0.221$), respectively; squares: the experimental data, after Katano 
         {\it et al.}~\cite{Katano2014A}.}
\label{fig:lambda+temp}
\end{figure}

The electron-phonon coupling constant $\lambda$ and its evolution with pressure can be determined with use of the equation:
\begin{equation}\label{eq:a2F-lambda}
\lambda = 2\int_0^{\omega_{\mathsf{max}}} \alpha^2F(\omega) \frac{d\omega}{\omega}.
\end{equation}
The results are collected in Table~\ref{tab:lambda}, the distribution of $\lambda$ over frequencies is analyzed in Fig.~\ref{fig:lomega}, and 
the total value is plotted against pressure in Fig.~\ref{fig:lambda+temp}. First of all we see, that calculations predict monotonic increase of 
$\lambda$ with pressure. The obtained value of $\lambda(p=0)=0.515$ agrees very well with the value of 0.52, reported in 
Ref.~\cite{Subedi2009A}. Analyzing  Fig.~\ref{fig:lomega} one can see, that due to the separation of $\alpha^2F(\omega)$ into three main parts,
$\lambda(\omega)$ is a three-step-like function, increasing rapidly in ranges 2-5~THz, 8-12~THz, and around 15 THz. The 40-THz high-frequency C-C mode contribute to $\lambda$ in less than $2\%$, whereas the generally monotonic increase of $\lambda(\omega)$ in the range 2-20 THz shows that most of the remaining phonon modes give important contribution to $\lambda$. 
For the applied pressures, the largest increase of $\lambda(\omega)$, comparing to $p = 0$~GPa, is seen between the first and second 'step' 
in Fig.~\ref{fig:lomega}, i.e. for the 8-12~THz modes. In the investigated pressures range, the $\lambda(p)$ increase linearly with a ratio of 
$\simeq 0.004$ per GPa, reaching 0.58 for $p = 15$~GPa.

\begin{table}[b]
\caption{Calculated electron-phonon coupling constant ($\lambda$) and the critical temperature ($T_c$ calc.) from the Allen-Dynes formula as a function of pressure. We have assumed $\mu^{\star}=0.13$. Experimental values of the critical temperature ($T_c$ expt.) taken from Ref.~\cite{Katano2014A}.}
\label{tab:lambda}
\begin{center}
\begin{ruledtabular}
\begin{tabular}{ c | c c c }
$p$ (GPa) & $\lambda$ & $T_c$ calc. (K) & $T_c$ expt. (K)\\
\hline
0	&	0.515&	2.82&	2.8	\\
2	&	0.522&	3.06&	3.4	\\
4	&	0.527&	3.22&	3.8	\\
5.5	&	0.533&	3.40&	3.0\footnotemark[1]	\\
 7      &	0.540&	3.60&	$\sim$ 0	\\
10	&	0.554&	4.03&	--	\\
15	&	0.581&	4.76&	--	\\
\end{tabular}
\footnotetext[1]{Interpolated between points for 4 and 6 GPa on the phase diagram from Ref.~\cite{Katano2014A}}
\end{ruledtabular}
\end{center}
\end{table}

Having $\alpha^2F(\omega)$, the superconducting critical temperature may be calculated in two ways. First, using the Allen-Dynes approximated formula~(\ref{eq:tc}), which is done now, and second, using the full Eliashberg gap equations, which is discussed in Sec.~\ref{sec:eliash}. The values of $T_c$, computed using the most often used Coulomb pseudopotential value $\mu^{\star} = 0.13$, are presented in Table~\ref{tab:lambda} and plotted versus pressure in the lower panel of Fig.~\ref{fig:lambda+temp}. Experimental values, as reported by Katano {\it et al.}~\cite{Katano2014A} are also given for  comparison. For the ambient conditions ($p=0$~GPa), the calculated $T_c$ agrees perfectly with the experimental value, which was also reported in Ref.~\cite{Subedi2009A}. Although this remarkable agreement depends on the value of $\mu^{\star}$, this may suggest that superconductivity in LaNiC$_2$ (at least without external pressure) is typical and well described in standard analytical
approach. As we will see in next Section, the full Eliashberg formalism, even at $p=0$~GPa, shows that it is not the case, and LaNiC$_2$ is an unusual superconductor in many aspects.  

When the external pressure is applied, superconducting $T_c$ increases monotonically, as a result of increase of $\lambda$, despite the small turnover in the behavior of the logarithmic average  $\langle \omega_{\rm log}^{\alpha^2F} \rangle$, reported in Table \ref{tab:freq} 
(it increases until 7~GPa pressure and then falls down by a small value).


This increase in $T_c$ qualitatively agrees with the experimental findings until $~\sim 3$-$4$~GPa, where even larger increase in $T_c$ was observed~\cite{Katano2014A}. But more interestingly, computed $\lambda$ and $T_c$ increase even above 4~GPa, where in experiment $T_c$ starts decreasing, and LaNiC$_2$ was reported not to superconduct above $\sim 7$~GPa~\cite{Katano2014A}. 

These results allow to draw at least two conclusions: 
(i) the experimentally observed increase in $T_c$ until 4~GPa may be explained by the increase in the electron-phonon coupling; 
(ii) the decrease of $T_c$ and the absence of superconductivity at larger pressures is not a typical effect of crystal 
lattice stiffening, and is not related to the decrease of $\lambda$, since the computed $\lambda$ continuously increases with pressure.
The second observation supports the hypothesis of the formation of a new electronic phase in LaNiC$_2$, however the driving 
force for this new phase hasn't been established yet. 
Note that, if to take into account only the electron-phonon interaction 
and the phase diagram of LaNiC$_2$ obtained by Katano {\it et al.}~\cite{Katano2014A}, in a natural way one can associate the new phase with the charge density waves. The stability of each of these states (superconductivity and CDW) depends sensitively on the temperature, the value of the electron-phonon coupling constant at the external pressure, and the details of electronic bands close to the Fermi surface \cite{Balseiro1979A, Gabovich2002A}. 
Therefore generally one could expect in the considered system occurrence of the pure metallic state, the pure superconducting state, the pure CDW, both metallic and semiconducting, and the coexistence of superconductivity and CDW.

%
\section{Eliashberg formalism\label{sec:eliash}}
%

%
\subsection{Eliashberg equations}

In this Section, we verify whether the full Eliashberg formalism can change above-mentioned conclusions and to what extent the superconductivity 
in LaNiC$_2$ differs from the BCS-like.
The Eliashberg equations on the imaginary axis ($i=\sqrt{-1}$) can be written in the following form: 
\begin{equation}
\label{r1}
\phi_{m}=\frac{\pi}{\beta}\sum_{n=-M}^{M}
\frac{\Lambda\left(i\omega_{m}-i\omega_{n}\right)-\mu^{\star}\theta\left(\omega_{c}-|\omega_{n}|\right)}
{\sqrt{\omega_n^2Z^{2}_{n}+\phi^{2}_{n}}}\phi_{n},
\end{equation}
\begin{equation}
\label{r2}
Z_{m}=1+\frac{1}{\omega_{m}}\frac{\pi}{\beta}\sum_{n=-M}^{M}
\frac{\Lambda\left(i\omega_{m}-i\omega_{n}\right)}{\sqrt{\omega_n^2Z^{2}_{n}+\phi^{2}_{n}}}\omega_{n}Z_{n},
\end{equation}
where $\phi_{m}=\phi\left(i\omega_{m}\right)$ represents the order parameter function and $Z_{m}=Z\left(i\omega_{m}\right)$ is the wave function renormalization factor. The physical value of the order parameter is given by the expression: $\Delta_{m}=\phi_{m}/Z_{m}$. The quantity $\omega_{m}$ denotes the Matsubara frequency: $\omega_{m}=\left(\pi /\beta\right)\left(2m-1\right)$, where $\beta=\left(k_{B}T\right)^{-1}$ is the inverse temperature. The pairing kernel of the electron-phonon interaction is traditionally written with the help of: 
$\Lambda\left(z\right)= 2\int_0^{\omega_{\mathsf{max}}}d\omega\frac{\omega}{\omega^{2}-z^{2}}\alpha^{2}F\left(\omega\right)$.
The symbol $\theta$ denotes the Heaviside function and $\omega_{c}$ represents the cut-off frequency: $\omega_{c}=10\omega_{\mathsf{max}}$.

The set of Eliashberg equations has one free parameter ($\mu^{\star}$), which models the depairing correlations. It is usually called the Coulomb pseudopotential, which is justified when $\mu^{\star}\leq 0.2$, since then $\mu^{\star}$ is associated with the renormalized Coulomb repulsion existing between the electrons \cite{Morel1962A}. For the values much higher than 0.2, the quantity $\mu^{\star}$ cannot result from the electron-electron repulsion only, and it hides in itself the additional physical effects which compete with the superconductivity, and which are not included explicitly in the classic Eliashberg formalism. However, the name of the Coulomb pseudopotential is used even in such cases.

In our studies of the termodynamic properties of the superconducting phase, for all the pressures, $\mu^{\star}(p)$ is chosen in such a way that the critical temperature, determined in the framework of the Eliashberg formalism, accurately reproduces the experimental value of $T_c$.  {Additionally, critical temperature is computed in two more ways: using the most commonly used $\mu^{\star} = 0.13$, as well as keeping the $\mu^{\star} = 0.221$, determined for $p=0$, constant for other pressures.} $T_c$ obtained in these ways is compared to that predicted by the Allen-Dynes formula. 

The Eliashberg equations were solved numerically for 2201 Matsubara frequencies. We used the numerical procedures described and tested for various materials in Refs.~\cite{Szczesniak2006A, Durajski2014A, Szczesniak2014B, Szczesniak2014C, Szczesniak2014D, Durajski2015A, Durajski2015B}. 
In the considered case, the stable solutions for the functions $\phi_{m}$ and $Z_{m}$ were obtained for the temperatures higher or equal to $T_{0}=1.8$~K.

The accurate values of the superconducting energy gap and the electron effective mass should be determined on the basis of the solutions of the Eliashberg equations defined on the real axis. In the case of the classical Eliashberg formalism, it is the most convenient to analytically continue the functions $\phi_{m}\rightarrow\phi\left(\omega\right)$ and $Z_{m}\rightarrow Z\left(\omega\right)$ with the help of the equations \cite{Marsiglio1988A}: 
\begin{widetext}
\begin{eqnarray}
\label{r3}
\phi\left(\omega+i\delta\right)&=&
                                  \frac{\pi}{\beta}\sum_{m=-M}^{M}
                                  \left[\Lambda\left(\omega-i\omega_{m}\right)-\mu^{\star}\theta\left(\omega_{c}-|\omega_{m}|\right)\right]
                                  \frac{\phi_{m}}
                                  {\sqrt{\omega_m^2Z^{2}_{m}+\phi^{2}_{m}}}\\ \nonumber
                              &+& i\pi\int_{0}^{+\infty}d\omega^{'}\alpha^{2}F\left(\omega^{'}\right)
                                  \left[\left[N\left(\omega^{'}\right)+f\left(\omega^{'}-\omega\right)\right]
                                  \frac{\phi\left(\omega-\omega^{'}+i\delta\right)}
                                  {\sqrt{\left(\omega-\omega^{'}\right)^{2}Z^{2}\left(\omega-\omega^{'}+i\delta\right)
                                  -\phi^{2}\left(\omega-\omega^{'}+i\delta\right)}}\right]\\ \nonumber
                              &+& i\pi\int_{0}^{+\infty}d\omega^{'}\alpha^{2}F\left(\omega^{'}\right)
                                  \left[\left[N\left(\omega^{'}\right)+f\left(\omega^{'}+\omega\right)\right]
                                  \frac{\phi\left(\omega+\omega^{'}+i\delta\right)}
                                  {\sqrt{\left(\omega+\omega^{'}\right)^{2}Z^{2}\left(\omega+\omega^{'}+i\delta\right)
                                  -\phi^{2}\left(\omega+\omega^{'}+i\delta\right)}}\right],
\end{eqnarray}
and
\begin{eqnarray}
\label{r4}
Z\left(\omega+i\delta\right)&=&
                                  1+\frac{i}{\omega}\frac{\pi}{\beta}\sum_{m=-M}^{M}
                                  \Lambda\left(\omega-i\omega_{m}\right)
                                  \frac{\omega_{m}Z_{m}}
                                  {\sqrt{\omega_m^2Z^{2}_{m}+\phi^{2}_{m}}}\\ \nonumber
                              &+&\frac{i\pi}{\omega}\int_{0}^{+\infty}d\omega^{'}\alpha^{2}F\left(\omega^{'}\right)
                                  \left[\left[N\left(\omega^{'}\right)+f\left(\omega^{'}-\omega\right)\right]
                                  \frac{\left(\omega-\omega^{'}\right)Z\left(\omega-\omega^{'}+i\delta\right)}
                                  {\sqrt{\left(\omega-\omega^{'}\right)^{2}Z^{2}\left(\omega-\omega^{'}+i\delta\right)
                                  -\phi^{2}\left(\omega-\omega^{'}+i\delta\right)}}\right]\\ \nonumber
                              &+&\frac{i\pi}{\omega}\int_{0}^{+\infty}d\omega^{'}\alpha^{2}F\left(\omega^{'}\right)
                                  \left[\left[N\left(\omega^{'}\right)+f\left(\omega^{'}+\omega\right)\right]
                                  \frac{\left(\omega+\omega^{'}\right)Z\left(\omega+\omega^{'}+i\delta\right)}
                                  {\sqrt{\left(\omega+\omega^{'}\right)^{2}Z^{2}\left(\omega+\omega^{'}+i\delta\right)
                                  -\phi^{2}\left(\omega+\omega^{'}+i\delta\right)}}\right]. 
\end{eqnarray}
\end{widetext}
The symbols $N\left(\omega\right)$ and $f\left(\omega\right)$ are the Bose-Einstein and the Fermi-Dirac functions, respectively.
 
%
\subsection{Numerical results}
\begin{figure}[t]
\centering
\includegraphics[width=0.97\columnwidth]{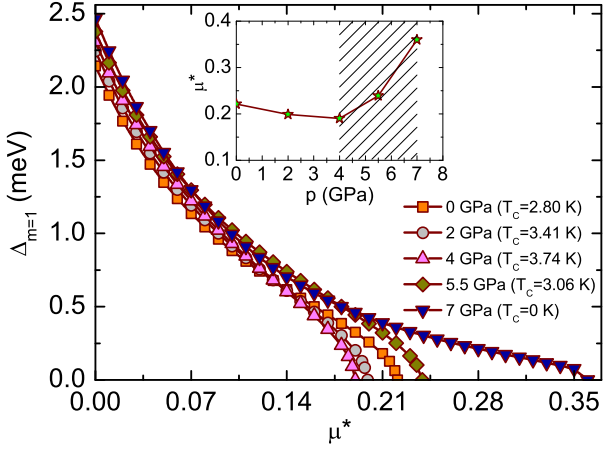}
\caption{The dependence of the maximum value of the order parameter on $\mu^{\star}$ ($T=T_c$ was assumed). The values of the parameter 
         $\mu^{\star}$ as a function of the pressure have been plotted in the insert. The hatched area corresponds to the pressure range, 
         wherein the superconducting state coexists with the high-temperature phase.}
\label{Fig01(III)}
\end{figure}

We start the discussion of the result with $p = 0$~GPa case. The Eliashberg equations with $\mu^{\star} = 0.13$ predict much larger value of the critical temperature $T_c = 4.8$~K, comparing to the Allen-Dynes formula ($T_c = 2.8$~K). Additionally, the Eliashberg equations with the $\mu^{\star} = 0.13$ value were solved for $p=7$~GPa and similar behavior was found: the critical temperature $T_c = 5.8$~K is considerably lager than the Allen-Dynes formula predicts (3.6 K) (see, also Fig.~\ref{fig:lambda+temp}). {The disagreement of $T_c$ between the Eliashberg equations and the Allen-Dynes formula stems from the fact that the Coulomb pseudopotential in the Eliashberg equations depends on the cut-off frequency} {$\omega_c$ \cite{Carbotte1990A}. Thus, if we want to discuss the validity of the Allen-Dynes formula, different $\mu^\star$ should be used in both theories. To correctly evaluate the thermodynamic properties in the framework of Eliashberg equations, the values of 
$\mu^{\star}$, which have 
to be applied to get $T_c$ consistent with experiment, have been calculated using the condition, that the superconducting energy gap vanishes at $T_c$: 
$\left[\Delta_{m={1}}\left(\mu^{\star}\right)\right]_{T=T_c}=0$. The explicit form of the function $\Delta_{m={1}}\left(\mu^{\star}\right)$ is presented in \fig{Fig01(III)}. Additionally, the insert shows the dependence of $\mu^{\star}$ on the pressure. It can be clearly seen that for $p=0$ GPa, we have obtained $\mu^{\star}=0.221$. 
Note that if we keep this value $\mu^{\star}=0.221$ constant with pressure, $T_c$ determined by the Eliashberg equations and the critical temperature from the Allen-Dynes formula for $\mu^{\star}=0.13$ are almost identical (see Fig.~\ref{fig:lambda+temp}), and to reproduce the experimental $T_c$, small variations in the $\mu^{\star}(p)$ function are needed. Nevertheless, for the whole pressure range up to 4 GPa, in which the sole superconducting state was experimentally found to exist, $\mu^{\star}\sim 0.2$.} Above $p = 4$~GPa, where $T_c$ decreases and coexistence of the superconducting state and the high-temperature electronic phase is expected~\cite{Katano2014A}, $\mu^{\star}$ has to rapidly grow with the increasing pressure, and the superconducting phase vanishes for $\mu^{\star}=0.36$ at $p\sim 7$~GPa, 
(see also \tab{t01(III)}). 

\begin{figure}
\centering
\includegraphics[width=0.97\columnwidth]{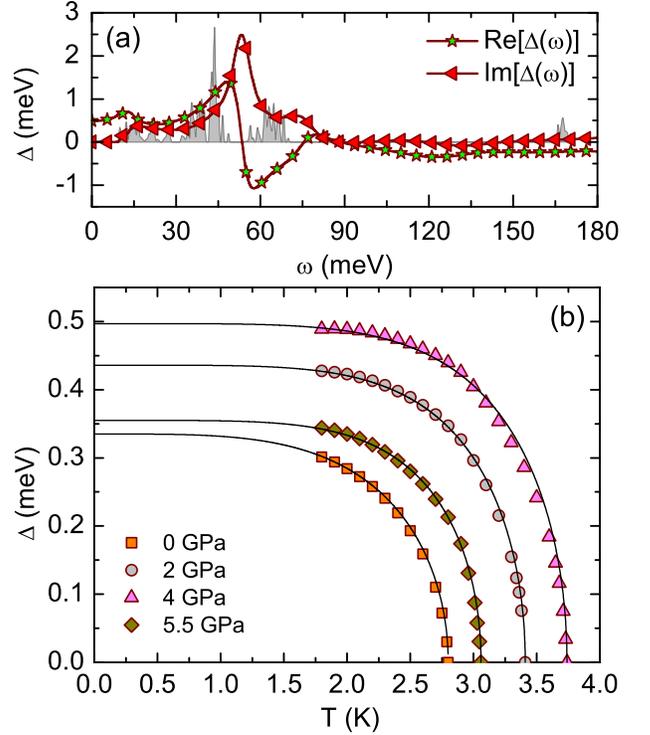}
\caption{(a) The form of the order parameter on the real axis obtained for $p=4$~GPa and $T=T_{0}$. 
         The rescaled Eliashberg function is plotted in the background: $2\alpha^{2}F\left(\omega\right)$. 
         (b) The order parameter as a function of temperature. Symbols represent the numerical results. Lines have been obtained 
         by using the formula (\ref{r6}) with exponent $\Gamma$ equal to 3.8 for $p=0$~GPa and 5 for higher pressures, instead of 3 predicted by the BCS 
         model \cite{Eschrig2001A}.}
\label{Fig02(III)}
\end{figure}

\begin{figure}
\centering
\includegraphics[width=0.97\columnwidth]{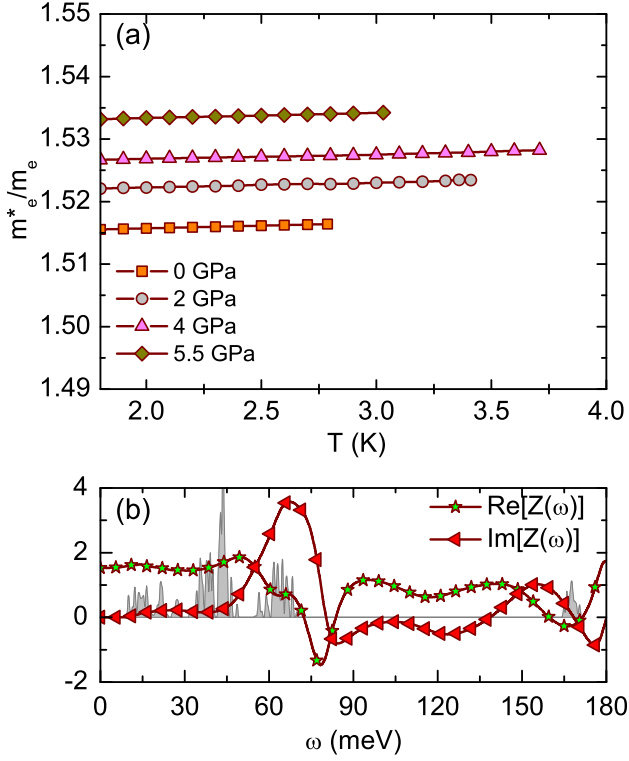}
\caption{(a) The electron effective mass as a function of the temperature. 
         (b) The wave function renormalization factor on the real axis for $p=4$~GPa and $T=1.8$~K. 
         The rescaled Eliashberg function $4\alpha^{2}F\left(\omega\right)$ is plotted in the background.}
\label{Fig03(III)}
\end{figure}
\begin{table}[b]
\centering
\caption{The values of the thermodynamic parameters obtained for the LaNiC$_2$ superconductor in the framework of the Eliashberg formalism.
$R_{\Delta}$ =$2\Delta\left(0\right)/k_BT_c$ and 
$R_{C}=\Delta C\left(T_c\right)/C^{N}\left(T_c\right)$.}
\label{t01(III)}
\begin{ruledtabular}
\begin{tabular}{llllll} 
$p$ (GPa)                                            & 0          & 2        & 4        & 5.5       & 7        \\ \hline
$\mu^{\star}$                                        & 0.221      & 0.199    & 0.19     & 0.239     & 0.36     \\
$\Delta\left(0\right)$ (meV)                         & 0.33       & 0.44     & 0.5      & 0.35      & -        \\
$R_{\Delta}$                                         & 2.74       & 3.00     & 3.05     & 2.70      & -        \\
$R_{C}$                                              & 1.14       & 1.68     & 2.05     & 1.2       & -        \\ 
\end{tabular}
\end{ruledtabular} 
\end{table} 

The accurate value of the order parameter for the given temperature was calculated on the basis of the equation~\cite{Carbotte1990A}:   
\begin{equation}
\label{r5}
\Delta\left(T\right)={\rm Re}\left[\Delta\left(\omega=\Delta\left(T\right)\right)\right].
\end{equation}
Results for the temperatures between the lowest $T_{0} = 1.8$~K, considered in our calculations, and $T_c$ are plotted as a function of pressure in~\fig{Fig02(III)}(b). 

The exemplary form of the order parameter on the real axis is presented in \fig{Fig02(III)}(a). Note that the values of the function 
$\Delta\left(\omega\right)$ are complex, whereas for the low frequencies the imaginary part of the order parameter is equal to zero. This means the infinite lifetime of the Cooper pairs due to the absence of the damping processes~\cite{Varelogiannis1997A}. Furthermore, the strong correlation between the course of the function $\Delta\left(\omega\right)$ and the shape of the Eliashberg function 
(the distribution of the maxima and the minima of these functions) draws the special attention. This is the characteristic feature of 
the solutions of the Eliashberg equations on the real axis. 
In the case of the frequencies higher than $\omega_{\mathsf{max}}$, the shape of 
the order parameter function becomes smoother. Additionally: 
${\rm \lim_{\omega\rightarrow +\infty}}{\rm Re}\left[\Delta\left(\omega\right)\right]=r$ and 
${\rm \lim_{\omega\rightarrow +\infty}}{\rm Im}\left[\Delta\left(\omega\right)\right]=0$, where $r$ is a negative real 
number~\cite{Zheng2007A}. 

The full dependence of the order parameter on temperature is shown in~\fig{Fig02(III)}(b). The numerical results can be described rather well using a simple analytic formula: 
\begin{equation}
\label{r6}
\Delta\left(T\right)=\Delta\left(0\right)\sqrt{1-\left(\frac{T}{T_c}\right)^{\Gamma}}, 
\end{equation}
where the exponent $\Gamma$ takes the value $3.8\pm0.3$ for 0~GPa and $5\pm0.3$ for other pressures. We notice that the BCS model predicts $\Gamma\simeq3$ \cite{Eschrig2001A}. With the help of Eq.~\ref{r6} the zero-temperature value of the superconducting order parameter $\Delta(0)$ have been extrapolated and are collected in~\tab{t01(III)}. The errors of the extrapolations are about 5\% for $p=0$ GPa and 2\% for the other pressures.

The values of the dimensionless ratio $R_{\Delta}=2\Delta\left(0\right)/k_{B}T_c$ are also presented in~\tab{t01(III)}. 
For $p=0$~GPa computed $R_{\Delta}=2.74$ lies between the two latest experimental results, equal to 2.5~\cite{Chen2013A} and 2.9~\cite{Hirose2012A} (see, \tab{tab:1}), and is considerably lower, than the BCS value of 3.53. In the full studied pressures range, the values of $R_{\Delta}$ are relevantly lower than the BCS one, and follow the $T_c$ trend, i.e. increase until 4 GPa, and then decrease.

With explicit solutions of the Eliashberg equations on the real axis, it is possible to precisely determine the temperature dependence of 
the electron effective mass $m^{\star}$ (the electron band mass renormalized by the electron-phonon interaction). To do this, the following formula should be used: 
\begin{equation}
\label{r7}
m^{\star}_{e}={\rm Re}\left[Z\left(\omega=0\right)\right]m_{e}, 
\end{equation}
where $m_{e}$ represents the electron band mass. In~\fig{Fig03(III)}(a) the $m^{\star}(T,p)$ curves are presented. We observe, that the electron effective mass increases with pressure, as expected due to the increase of $\lambda$, and weakly dependents on temperature, reaching a largest value for $T=T_c$, where $\left[m^{\star}_{e}\right]_{\rm max}\simeq\left(1+\lambda\right)m_{e}$. 
As an illustrative example, the wave function renormalization factor is shown in \fig{Fig03(III)}(b) for $p = 4$ GPa. Similarly as it was for the order parameter, the correlation between the shape of the functions, $Z\left(\omega\right)$ and $\alpha^2F(\omega)$, is visible.   
\begin{figure}
\centering
\includegraphics[width=0.97\columnwidth]{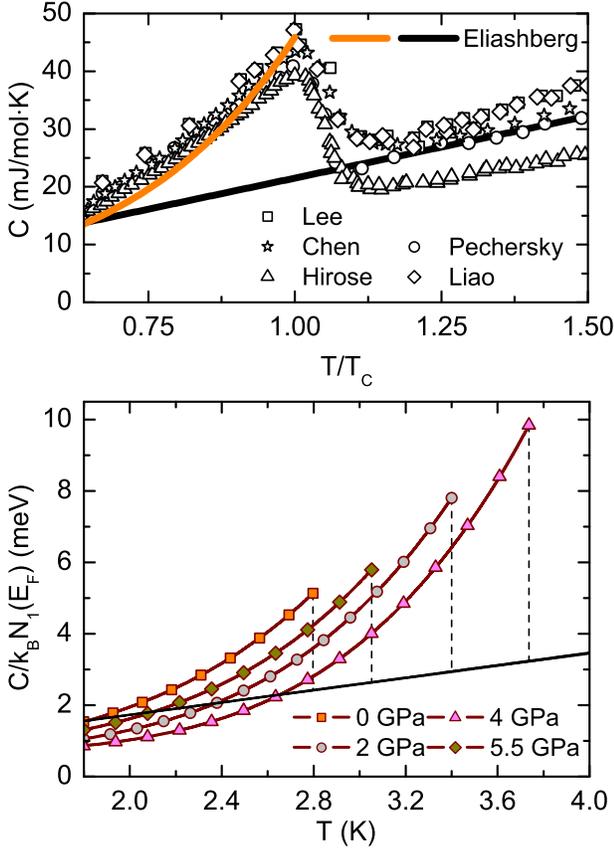}
\caption{Upper panel: The computed specific heat of the superconducting state as a function of the reduced temperature for $p=0$~GPa, compared with the experimental data: Lee {\it et al.}~\cite{Lee1996A}, Chen {\it et al.}~\cite{Chen2013A}, Hirose {\it et al.}~\cite{Hirose2012A}, 
         Pecharsky {\it et al.}~\cite{Pecharsky1998A} and Liao {\it et al.}~\cite{Liao2009A}.
         Lower panel: reduced specific heat of the superconducting state (lines with symbols) as a function of temperature and pressure. 
         Additionally, the linear contribution to the specific heat of the normal state (black solid line) has been plotted. }
\label{Fig05(III)}
\end{figure}

The condensation energy, i.e. the difference in the free energy between the superconducting and normal states has been calculated on the basis of the solutions of the Eliashberg equations on the imaginary axis~\cite{Carbotte1990A}:
\begin{eqnarray}
\label{r9}
\frac{\Delta F}{N_{1}\left(E_{F}\right)}&=&-\frac{2\pi}{\beta}\sum_{n=1}^{M}
\left(\sqrt{\omega^{2}_{n}+\Delta^{2}_{n}}- \left|\omega_{n}\right|\right)\\ \nonumber
&\times&\left(Z^{S}_{n}-Z^{N}_{n}\frac{\left|\omega_{n}\right|}
{\sqrt{\omega^{2}_{n}+\Delta^{2}_{n}}}\right),
\end{eqnarray}  
where $N_1\left(E_{F}\right)$ represents the value of electron density of states on the Fermi level per spin.
The symbols $Z^{S}_{n}$ and $Z^{N}_{n}$ denote the wave function renormalization factor for the superconducting state and the normal state, respectively.  The condensation energy determines the electronic specific heat difference between the superconducting and the normal states:
\begin{equation}
\label{r11}
\frac{\Delta C\left(T\right)}{k_{B}N_1\left(E_{F}\right)}=-\frac{1}{\beta}\frac{d^{2}\left[\Delta F/N_1\left(E_{F}\right)\right]}
{d\left(k_{B}T\right)^{2}},
\end{equation}
where the specific heat of the normal state is given by the formula: $C^{N}=\gamma{T}$, and the Sommerfeld constant has the form: 
$\gamma=\frac{2}{3}\pi^{2}k_{B}^{2}N_1\left(E_{F}\right)\left(1+\lambda\right)$. 

Specific heat has been plotted in \fig{Fig05(III)}, where the upper panel shows the $p=0$ GPa results, compared with the rich collection of the experimental data, and the lower panel shows the evolution of the specific heat (in reduced units) with pressure. 
The upper panel shows, that within the range of the measurement precision, the Eliashberg formalism correctly reproduces the values and the temperature dependence of the specific heat for the superconducting state and the normal state. 

Then, the characteristic dimensionless ratio $R_{C}=\Delta C\left(T_c\right)/C^{N}\left(T_c\right)$ is calculated. The results are presented in~\tab{t01(III)}, and the value for ambient pressure is $R_{C} = 1.14$, which rather well reproduces the experimental results $1.05$-$1.20$ (see, \tab{tab:1}). We notice, that the obtained values of $R_{C}(p)$ are pressure-dependent and follow the same trend as $R_{\Delta}$ and $T_c$, increasing until 4 GPa. The deviation from the predictions of the BCS theory ($\left[R_{C}\right]_{\rm BCS}=1.43$)~\cite{Bardeen1957A, Bardeen1957B} again shows the non-BCS type of the superconductivity of LaNiC$_2$, but within the s-wave Eliashberg formalism. 
\begin{figure}
\centering
\includegraphics[width=0.97\columnwidth]{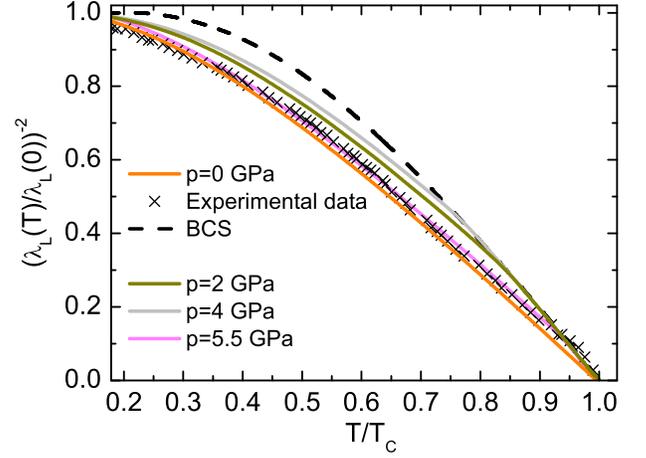}
\caption{The normalized London penetration depth as a function of the reduced temperature. The experimental results for the normal pressure 
         are taken from Ref.~\cite{Chen2013A}; the value of $\lambda_L\left(0\right)$ is $3940$ ${\rm \AA}$.}
\label{Fig06(III)}
\end{figure}

In the last step, the temperature dependence of the London penetration depth ($\lambda_{L}$) has been calculated:
\begin{eqnarray}
\label{r12}
\frac{1}{e^2v^2_F N_{1}(E_{F})\lambda^2_L\left(T\right)}=\frac{4}{3}\frac{\pi}{\beta}\sum_{n=1}^{M}
\frac{\Delta^{2}_{n}}{Z^{S}_{n}\left[ \omega^{2}_{n}+\Delta^{2}_{n}\right]^{3/2}},
\end{eqnarray}
where $e$ is the electron charge and $v_{F}$ is the Fermi velocity~\cite{Carbotte1990A}. 
The normalized $\lambda^{-2}_L$ as a function of the reduced temperature is presented in~\fig{Fig06(III)}. For $p=0$~GPa, the theoretical curve very precisely reproduces the latest experimental results~\cite{Chen2013A}, and both results differ significantly from the classical BCS theory predictions. Thus, the deviation of $\lambda_L(T)$ from the BCS curve in LaNiC$_2$ may result {either from the two s-wave gaps, as proposed in Ref~\cite{Chen2013A}, or from the retardation effects included in the Eliashberg formalism, as shown here, and does not prejudge the existence of multiple gaps or unconventional pairing symmetry. }
The intermediate results between the curve for $p=0$~GPa and the BCS curve are predicted for the higher pressures.  

%
\section{Summary and Conclusions}

The results presented in Sections~\ref{sec:abinitio} and \ref{sec:eliash} show that the LaNiC$_2$ compound exhibits unconventional, non-BCS superconducting properties, but with the electron-phonon interaction being most likely the pairing mechanism. 
The magnitude of the electron-phonon coupling constant $\lambda \simeq 0.5$-$0.6$ confirms the weak to intermediate coupling regime of the interaction. In spite of this, thermodynamics of the superconducting phase is not BCS-like.
First, under ambient pressure, the superconductivity may be accurately described using the s-wave Eliashberg formalism, but only if the Coulomb pseudopotential parameter is set to $\mu^{\star} = 0.22$. Such a larger value of $\mu^{\star}$ itself is not something exceptional, similar values were needed to reproduce experimental $T_c$ e.g. for V (0.30), Nb (0.21)~\cite{Savrasov1996}, Nb$_3$Ge (0.24)~\cite{Carbotte1990A}, MgCNi$_3$ (0.29)~\cite{Szczesniak2015A} or superconducting high entropy alloy TaNbHfZrTi (0.25)~\cite{Jasiewicz2016}.
{However, the agreement between the Allen-Dynes value of $T_c$ at $p=0$~GPa and experiment has been obtained for the typical $\mu^{\star}=0.13$}. 

The thermodynamic properties of the superconducting state of LaNiC$_2$ become non-BCS-like, even if the s-wave symmetry of the order parameter is assumed, and the values of the characteristic parameters $R_{\Delta} =2.74$ and $R_{C} = 1.14$ significantly differ from these, predicted by the BCS theory (3.53 and 1.43, respectively). If $\mu^{\star} = 0.22$ is taken, not only the superconducting critical temperature is captured, but also the experimental temperature dependence of the London penetration depth $\lambda_L(T)$, which is non-BCS-like, is accurately reproduced.
Thus, in view of our results, the frequently used argument, that $\lambda_L(T)$ does not follow the BCS curve, shouldn't be treated as definite confirmation for the two-gap or the nodal-gap superconducting state. 
Also, the specific heat in the superconducting state seems to be well described within the Eliashberg formalism, but here the dispersion between the different experimental results is relatively large, which make this conclusion less firm. 
The formalism used here does not allow to analyze the triplet pairing, thus we are not able to discuss the non-unitary triplet pairing suggested for LaNiC$_2$ in Refs.~\cite{Hillier2009A, Quintanilla2010A} nor explain the existence of small magnetic fields, as found recently in Ref.~\cite{Sumiyama2015A}.

Under external pressure, the first principles computations do not show any spectacular changes in the electronic structure of the system. We observe a gradual increase of the Ni-$3d$ DOS at $E_F$, but the absolute values are too low to induce an ordered magnetic state. On the other hand, the electron-phonon coupling is enhanced by the pressure, and $\lambda$ increases from 0.515 (0~GPa) via 0.54 (7~GPa) to 0.58 at 15~GPa. This correlates with the experimentally observed increased $T_c$ of LaNiC$_2$ in the pressure range 0-4~GPa, and the $s$-wave Eliashberg formalism reproduces quantitatively the experimental $T_c$ when $\mu^{\star} \simeq 0.20$ is taken. 
For the larger pressures, when in experiment the decrease of $T_c$ is observed, and computations predict the continuous increase of 
$\lambda$, the Eliashberg formalism is not able to explain the observed trends, if the $\mu^{\star}$ value is kept at the previous level of 
$\mu^{\star}\sim 0.2$. That strongly supports the explanation, suggested by Katano {\it et al.}~\cite{Katano2014A}, that a new electronic phase is induced by the pressure in the LaNiC$_2$ compound, and is competing with the superconductivity, leading to disappearance of the superconducting state above $7$-$8$~GPa.
One could speculate, that the origin of this new phase and internal small magnetic fields, detected at $p=0$ GPa, could be the same, and that already at ambient pressure there is some competition between superconductivity and the ,,second'' phase. This could well explain the enhanced value of the Coulomb pseudopotential parameter $\mu^{\star} \simeq 0.2$, which was needed to reproduce $T_c$ basing on Eliashberg formalism. The muon spin relaxation $\mu$SR or magnetization measurements under pressure could shed more light on this issue and help to verify whether the internal magnetic fields, reported in Refs.~\cite{Hillier2009A,Sumiyama2015A}, are enhanced or quenched by the pressure, suggesting whether they compete or cooperate with the superconductivity. {Additionally, such measurements could be performed for the electron-doped samples, since, as we mentioned, electron doping and external pressure result in similar changes in the electronic structure and critical temperature of LaNiC$_2$.}

{Although Eliashberg formalism successfully explained a number of experimental results, we have to note, that in the presented analysis of the superconducting phase, spin-orbit coupling and Fermi surface splitting was not taken into account, and single-gap s-wave symmetry was assumed. We may speculate, that the Fermi surface splitting could result in reduction of $T_c$, and for this reason in our studies increased value of $\mu^{\star}$ was necessary to obtain computed $T_c$ equal to the experimental one. To resign from this assumption, but still considering singlet pairing, one would have to calculate the anisotropic, momentum-dependent Eliashberg coupling functions $\alpha^2F({\bf k},{\bf k'}, \omega$), taking into account spin-orbit interaction and Fermi surface splitting. Next, corresponding set of anisotropic Eliashberg equations would have to be solved, similarly to what was done for MgB$_2$~\cite{Choi2002A,Choi2002B}, but extended to include SOC. 
Since such calculations are beyond the scope of present work, further efforts are required to unambiguously conclude on the actual pairing symmetry in LaNiC$_2$, especially that some of the experimental results seem to contradict each other (see, Introduction).
However, as the effect of SOC on the electronic density of states is negligible, we do not expect a significant effect of SOC on either an overall magnitude of the electron-phonon interaction, or on the integrated Eliashberg function $\alpha^2F(\omega)$. Thus, our observations of increased strength of the electron-phonon coupling under pressure should remain valid, leaving the disappearance of superconductivity above 4 GPa in LaNiC$_2$ as related to different, than electron-phonon mechanism.}

%
\begin{acknowledgments}
BW was partially supported by the Polish Ministry of Science and Higher Education.
\end{acknowledgments}
\bibliography{LaNiC2}
\end{document}